\documentstyle[12pt]{article}
\begin{document}

\begin{center}
{\bf{Photons with half-integral spin as q-Fermions}}

\vspace{1.0cm}

R.Parthasarathy{\footnote{e-mail: sarathy@cmi.ac.in}} \\
Chennai Mathematical Institute \\
Chennai 603103, India \\
and \\
K.S.Viswanathan{\footnote{e-mail: kviswana@sfu.ca}} \\
Department of Physics \\
Simon Fraser University \\
Burnaby B.C, Canada \\
\end{center}

\vspace{1.0cm}

{\noindent{\bf{Abstract}}}

\vspace{0.5cm}

The recently discovered 'light (photons) with half-integral spin' is interpreted as q-Fermions proposed by 
us in 1991, as these q-Fermions satisfy q-deformed anti-commutation relations (pertaining to spin half) and 
have the property that more than one q-Fermion can occupy a given quantum state. In this article, in view of 
the recent discovery, we recall the construction of q-Fermions and give the statistical properties of 
q-Fermion gas, based on our preprint in 1992. 

\newpage 

Recently, Ballantine, Donegan and Eastham [1] reported the discovery of 'light (photons) with half-integral spin'.
This new form of light is interpreted as q-Fermions proposed by Parthasarathy and Viswanathan [2] in 1991. 
The q-Fermions satisfy q-deformed anti-commutation relations pertaining to spin half and have the novel 
property that more than one q-Fermion can occupy a given state. In the limit $q\rightarrow 1$, we recover 
ordinary fermions. With $q\neq 1$, the q-Fermion system could be the right description of the newly 
discovered light with half-integral spin. We [3] have given the q-deformed statistics of q-Fermions, 
obtaining the number density, chemical potential and the specific heat $C_V$. In view of the recent discovery,
we give the statistical properties of q-Fermion gas, essentially taken from the preprint [3]. 

\vspace{0.5cm}

The result in [1] shows, in two dimensions photons can have a half-integer total angular momentum, 
comprising of unequal mixture of spin and orbital contributions and demonstrate the half-integer 
quantization of this total angular momentum. They involve beams of light propagating in a particular 
direction when the full rotational symmetry is not present. The restricted symmetry leads to a new 
form of total angular momentum which has a half-integer; that is 'fermionic' spectrum. This 
half-integer spectrum shows that for light, reduced dimensionality allows for new form of 
quantization. The half-integer quantization , which they demonstrate through noise measurement, 
implies fermionic exchange statistics. A description of such photons could be provided by the 
q-Fermions and q-Fermion statistics. In our preprint [3], we have demonstrated that q-deformed 
version is not merely a mathematical construct but perhaps effectively takes into account the 
interactions. These interactions are contact interactions and either the {\it{surroundings}} or 
the system itself generate these interactions. It is in this sense that the restricted symmetry 
is effectively taken into account by the q-deformation. While the q-deformed Boson description 
maintains the commutation relations though q-deformed, to describe half-integer total angular 
momentum of the new photons, q-Fermion description is appropriate. Nevertheless, this 
q-Fermion describing the new photon cannot be a fundamental particle, but particle with 
interactions included, as if it is a quasi-particle.

\vspace{0.5cm}

{\noindent{\bf{q-Fermion Algebra - A brief Review}}}

\vspace{0.5cm}

A q-deformed non-trivial (in the sense that there is another version of q-fermion algebra [4]
which can be transformed to ordinary fermion algebra and hence trivial) fermion algebra proposed by us 
[2] is given by 
\begin{eqnarray}
ff^{\dagger}+\sqrt{q}f^{\dagger}f &=& q^{-\frac{N}{2}}, \nonumber \\
{[N,f]}&=&-f, \nonumber \\
{[N,f^{\dagger}]}&=&f^{\dagger}, \nonumber \\
f^2\ \neq \ 0 &;& (f^{\dagger})^2\ \neq \ 0,
\end{eqnarray}
where $N$ is the q-Fermion number operator $\neq f^{\dagger}f$. The orthonormal $n$ q-Fermion state is defined by 
\begin{eqnarray}
{|n\rangle}^F_q&=&\frac{1}{([n]^F_q!)^{\frac{1}{2}}}\ (f^{\dagger})^n|0\rangle,
\end{eqnarray}
where the vacuum $|0\rangle$ is defined as $f|0\rangle =0$ and 
\begin{eqnarray}
{[n]^F_q!}&=&[n]^F_q\ [n-1]^F_q\ \cdots [2]^F_q\ [1]^F_q, \nonumber \\
{[n]^F_q}& =& q^{-\frac{(n-1)}{2}}\sum_{k=0}^{(n-1)}(-q)^k\ =\ \frac{\sqrt{q}}{1+q}\left(q^{-\frac{n}{2}}- 
(-1)^nq^{\frac{n}{2}}\right).
\end{eqnarray}

\vspace{0.5cm}

This algebra describes q-fermions such that any number of q-fermions can occupy a given state for $0<q<1$. 
Here $q$ is taken to be real.  When $q=1$, all states other than the vacuum and one particle state collapse, 
thereby recovering Pauli principle. Subsequently, Viswanathan, Parthasarathy and Jagannathan [5]  have 
constructed coherent states for q-fermions. In it, we used transformed q-fermion operator as  
$f=q^{-\frac{N}{4}}F,\ f^{\dagger}=F^{\dagger}q^{-\frac{N}{4}}$, so that we find equivalent algebra 
\begin{eqnarray}
FF^{\dagger}+qF^{\dagger}F&=&1, \nonumber \\
FF^{\dagger}-F^{\dagger}F&=& (-q)^N, \nonumber \\
{[N,F]}&=&-F, \nonumber \\
{[N,F^{\dagger}]}&=&F^{\dagger}, \nonumber \\
F^2\ \neq \ 0 &;& (F^{\dagger})^2\ \neq \ 0, 
\end{eqnarray}
from which we see 
\begin{eqnarray}
FF^{\dagger}\ =\ [N+1]^f &;& F^{\dagger}F\ =\ [N]^f, 
\end{eqnarray}
where 
\begin{eqnarray}
{[n]^f}&=&\frac{1-(-q)^n}{1+q}. 
\end{eqnarray}
The superscript 'f' is used to indicate that we are dealing with q-fermion. Here $N$ is the number operator, 
$\neq \ F^{\dagger}F$. The Fock space $|n\rangle$ can be constructed as 
\begin{eqnarray}
F|n\rangle &=& ([n]^f)^{\frac{1}{2}}|n-1\rangle, \nonumber \\
F^{\dagger}|n\rangle &=&([n+1]^f)^{\frac{1}{2}}|n+1\rangle .
\end{eqnarray}
As before, the vacuum is defined by $F|0\rangle =0$. For $q\neq 1$, more than one q-Fermion can occupy a 
given state and {\it{only when $q=1$, all states other than $|0\rangle$ and $|1\rangle$ collapse, recovering 
Pauli principle. The above q-fermion algebra (either in terms of $f$ or $F$) is non-trivial as it cannot be 
transformed to ordinary fermion algebra.}} Using (6), we find 
\begin{eqnarray}
{[n+1]^f+q[n]^f}&=&1.
\end{eqnarray}

\vspace{4.0cm}

{\noindent{\bf{Many q-Fermion system - q-Fermion Gas}}}

\vspace{0.5cm}

As a model Hamiltonian for many q-fermion system, we take 
\begin{eqnarray}
H&=&\sum_k(E_k-\mu)N_k, 
\end{eqnarray}
where $E_k$ is the kinetic energy for q-fermion of momentum $k$. We assume that q-fermion operators with different 
momenta commute. We define the thermal average of $F^{\dagger}F$ as 
\begin{eqnarray}
\langle F^{\dagger}_kF_k\rangle &=&\frac{Tr(\exp(-\beta H)F^{\dagger}_kF_k)}{Tr\ \exp(-\beta H)}, 
\end{eqnarray}
where $\beta =\frac{1}{KT}$, $K$ the Boltzmann constant. Using the cyclic property of the trace and (4), it 
follows 
\begin{eqnarray}
\langle F^{\dagger}_kF_k\rangle &=& \exp (-\beta(E_k-\mu))\ \langle F_kF^{\dagger}_k\rangle.
\end{eqnarray}
Using (8), we find 
\begin{eqnarray}
\langle [N_k]^f\rangle &=& \frac{1}{\exp(\beta(E_k-\mu))+q},
\end{eqnarray}
{\it{which is the distribution function for q-fermions.}} It is noted that when $T\rightarrow 0$, for 
$E_k>\mu$, $\langle [N_k]^f\rangle \rightarrow 0$; for $E_k<\mu$, $\langle [N_k]^f\rangle \rightarrow 
\frac{1}{q}$ and for $E_k=\mu$, $\langle [N_k]^f\rangle \rightarrow \frac{1}{1+q}$. So upto 
$E_k={\mu}_0<\mu$, the levels are filled and empty when $E_k>\mu$. Further from (5) and (11), we see that 
\begin{eqnarray}
\frac{[N_k]^f}{[N_k+1]^f}&=&e^{-\beta(E_k-\mu)},
\end{eqnarray}
From (6), in the limit $q\rightarrow 1$, $[n]=n,\ [n+1]=1-n, \ n=0,1$, and so (13) gives the familiar 
Fermi-Dirac distribution when $q=1$.  

\vspace{0.5cm}

The distribution function (12) for many q-Fermion system derived by us in Ref.3 (in 1992) has been the subject of study by 
others, Narayana Swamy [6] in 2006, Algin and Senay [7] in 2012, Algin, Irk and Topcu [8] in 2015. The expression (12) can 
be solved for $N_k$ as 
\begin{eqnarray}
N_k&=&\frac{1}{|\ell nq|}|\ell n \left(\frac{|e^{{\eta}_k}-1|}{e^{{\eta}_k}+q}\right)|,
\end{eqnarray}
where ${\eta}_k=\beta(E_k-\mu)$, agreeing with [6], [7].  

\vspace{0.5cm}

The Hamiltonian (9) allows us to evaluate the number density $\rho$ and the internal energy $u$ for the 
q-Fermion gas, by going from discrete sum to integral. Following Feynman [9], we consider 
\begin{eqnarray}
I&=&\int_0^{\infty} \frac{g(E)dE}{\exp(\beta(E-\mu))+q}, 
\end{eqnarray}
where $g(E)=c\sqrt{E}$ for calculating the number density $\rho$ and $g(E)=cE\sqrt{E}$ for calculating the 
internal energy, $c$ a constant. Splitting the integral as 
\begin{eqnarray}
I&=&\frac{1}{q}\int_0^{\mu}g(E)dE-\frac{1}{q}\int_0^{\mu}\frac{g(E)dE}{1+q\exp(-\beta(E-\mu))}+
\int_{\mu}^{\infty}\frac{g(E)dE}{\exp(\beta(E-\mu))+q}, \nonumber \\
& &
\end{eqnarray}
which can be easily verified, setting $x=-\beta(E-\mu)$ in the second integral and $x=\beta(E-\mu)$ in the 
third integral, we find 
\begin{eqnarray}
I&=&\frac{1}{q}\int_0^{\mu} g(E)dE-\frac{1}{q\beta}\int_0^{\mu\beta}\frac{g(\mu-\frac{x}{\beta})dx}{1+qe^x}
+\frac{1}{\beta}\int_0^{\infty}\frac{g(\mu+\frac{x}{\beta})dx}{e^x+q}, \nonumber \\
& & 
\end{eqnarray}
For low enough temperatures $g(\mu \pm\frac{x}{\beta})\ \simeq \ g(\mu)\pm \frac{x}{\beta}g'(\mu)$ and 
$\int_0^{\mu\beta}\rightarrow \int_0^{\infty}$. Then
\begin{eqnarray}
I&=&\frac{1}{q}\int_0^{\mu}g(E)dE-\frac{g(\mu)}{q\beta}\int_0^{\infty}\frac{dx}{1+qe^x}+\frac{g'(\mu)}
{q{\beta}^2}\int_0^{\infty}\frac{xdx}{1+qe^x} \nonumber \\
&+&\frac{g(\mu)}{\beta}\int_0^{\infty}\frac{dx}{e^x+q}+\frac{g'(\mu)}{{\beta}^2}\int_0^{\infty}
\frac{xdx}{e^x+q}.
\end{eqnarray}
  
\vspace{0.5cm}

Now 
\begin{eqnarray}
\int_0^{\infty}\frac{dx}{1+qe^x}=\ell n\left(\frac{1+q}{q}\right)&;&\int_0^{\infty}\frac{dx}{e^x+q}
=\frac{1}{q}\ell n(1+q).
\end{eqnarray}
Collecting $g'(\mu)$ terms, we have 
\begin{eqnarray}
\frac{g'(\mu)}{{\beta}^2}\left(\frac{1}{q}\int_0^{\infty}\frac{xdx}{1+qe^x}+\int_0^{\infty}\frac{xdx}
{e^x+q}\right),
\end{eqnarray}
which can be evaluated as 
\begin{eqnarray}
\frac{g'(\mu)}{q{\beta}^2}\left(\frac{{\pi}^2}{6}+\frac{1}{2}(\ell nq)^2\right). 
\end{eqnarray}
Thus, 
\begin{eqnarray}
I&=&\frac{1}{q}\int_0^{\mu} g(E)dE+\frac{g(\mu)}{q\beta}\ell nq+\frac{g'(\mu)}{q{\beta}^2}
\left(\frac{{\pi}^2}{6}+\frac{1}{2}(\ell nq)^2\right). 
\end{eqnarray}

\vspace{0.5cm}

For finding the number density $\rho$, set $g(E)=c\sqrt{E}$ and then we find 
\begin{eqnarray}
\rho &=&\frac{2c}{3q}{\mu}^{\frac{3}{2}}+\frac{c\sqrt{\mu}}{q\beta}\ell nq+\frac{c}{2q{\beta}^2\sqrt{\mu}}
\left(\frac{{\pi}^2}{6}+\frac{1}{2}(\ell nq)^2\right).
\end{eqnarray}
From this, we see 
\begin{eqnarray}
{\rho}_{T=0}&=&\frac{2c}{3q}{{\mu}_0}^{\frac{3}{2}}.
\end{eqnarray}
Equating ${\rho}_{T=0}={\rho}_{T\neq 0}$ as required by number conservation, we find 
\begin{eqnarray}
{\mu}^{\frac{3}{2}}&=&{{\mu}_0}^{\frac{3}{2}}\left(1+\frac{3}{2\mu\beta}\ell nq+\frac{3}{4{\beta}^2{\mu}^2}
[\frac{{\pi}^2}{6}+\frac{1}{2}(\ell nq)^2]\right)^{-1}.
\end{eqnarray}

Here we approximate $\mu$ by ${\mu}_0$ in the $\left(......\right)^{-1}$ and then 
\begin{eqnarray}
\mu&\simeq&{\mu}_0\left(1-\frac{1}{{\mu}_0\beta}\ell nq-\frac{{\pi}^2}{12{{\mu}_0}^2{\beta}^2}+\frac{1}
{ {{\mu}_0}^2{\beta}^2}(\ell n q)^2\right).
\end{eqnarray}

\vspace{0.5cm}

The internal energy of q-fermion gas is evaluated by taking $g(E)=cE\sqrt{E}$ and we find 
\begin{eqnarray}
u&=& \frac{2c}{5q}{\mu}^{\frac{5}{2}}+\frac{c}{q\beta}{\mu}^{\frac{3}{2}}\ell nq+\frac{3c}{2q{\beta}^2}
\sqrt{\mu}\ [\frac{{\pi}^2}{6}+\frac{1}{2}(\ell nq)^2].
\end{eqnarray}
Using the expression for $\mu$ above and after some steps, the internal energy becomes 
\begin{eqnarray}
u&=&u_0+\gamma T^2, 
\end{eqnarray}
where 
\begin{eqnarray}
u_0&=&\frac{2c}{5q}{{\mu}_0}^{\frac{5}{2}}, \nonumber \\
\gamma &=& \frac{c\sqrt{{\mu}_0}K^2}{q}\ \frac{{\pi}^2}{6}+\frac{c\sqrt{{\mu}_0}K^2}{q}(\ell nq)^2, \nonumber \\
       &=&\frac{c\sqrt{{\mu}_0}K^2{\pi}^2}{6}\left(\frac{1}{q}+\frac{6}{{\pi}^2q}(\ell nq)^2\right).
\end{eqnarray}
The pre-factor in $\gamma$ is the Feynman's value and the correction is multiplicative parenthesis. Since 
$C_V=\frac{\partial u}{\partial T}$, we find for q-fermion gas 
\begin{eqnarray}
C_V&=&C_V^{Feynman}\left(\frac{1}{q}+\frac{6}{{\pi}^2q}(\ell nq)^2\right).
\end{eqnarray}
In the limit $q=1$, we recover Feynman's value. The results in (23), (25), (26), (27) and (30) were derived 
by us in our preprint [3].

\vspace{0.5cm}

By measuring $C_V$ for q-fermion gas, $q$ can be determined 
which can then be used to find $\rho$. These expressions can be applied to the 'newly discovered photon with 
half-integral spin', if the $\rho$, $C_V$ are measured for this system. Further, the expression for $N_k$ in (14) 
can be used to plot the q-deformed statistical distribution function for q-Fermion gas for various $T$ as a 
function of $\beta(E-\mu)$ for values of $q<1$. This plot is available from Fig.1 of [7] and can be used for the 
distribution function of the 'new light with half-integer spin'.  

\vspace{0.5cm}

{\noindent{\bf{References}}}

\vspace{0.5cm}

\begin{enumerate}
\item K.E. Ballantine, J.F. Donegan and P.R. Eastham, Science Advances. {\bf{2}}, Number 4 (2016) E1501748. April 2016.
\item R. Parthasarathy and K.S. Viswanathan, J.Phys.A:Math.Gen. {\bf{A24}} (1991) 613. 
\item R. Parthasarathy and K.S. Viswanathan, IMSc-Preprint - 92/57. December 15, 1992. 
\item S. Jing and J. Xu, J.Phys.A:Math.Gen. {\bf{A24}} (1991) L891. 
\item K.S. Viswanathan, R. Parthasarathy and R. Jagannathan, J.Phys.A:Math.Gen. {\bf{A25}} (1992) L335. 
\item P. Narayana Swamy, quant-ph/0509136.
\item A. Algin and M. Senay, Phys.Rev. {\bf{E85}} (2012) 041123.
\item A. Algin, D. Irk and G. Topcu, Phys.Rev. {\bf{E91}} (2015) 062131. 
\item R.P. Feynman, {\it{Statistical Mechanics - A set of Lectures}}, W.A. Benjamin, Inc. (1972).
\end{enumerate} 

\end{document}